\let\ifarxiv=\iftrue     % ARXIV VERSION
\pdfoutput=1

\ifarxiv

\documentclass[12pt,a4paper]{article}
\usepackage[a4paper,text={450pt,650pt},centering]{geometry}

\fi

\ifarxiv\else

\documentclass[11pt,a4paper]{article}
\usepackage{mathptmx}
\usepackage[a4paper,text={130mm,198mm}]{geometry}

\fi

\usepackage{amsmath,amssymb}
\usepackage[bookmarks=true,hyperfigures=true]{hyperref}
\usepackage{graphicx}
\usepackage[nosort]{cite}
\let\oldbfseries=\bfseries
\let\oldmdseries=\mdseries
\let\oldnormalfont=\normalfont
\renewcommand{\bfseries}{\oldbfseries\boldmath}
\renewcommand{\mdseries}{\oldmdseries\unboldmath}
\renewcommand{\normalfont}{\oldnormalfont\unboldmath}

\allowdisplaybreaks[3]

\numberwithin{equation}{section}

\usepackage[font=small,labelfont=bf,width=0.85\textwidth]{caption}

\providecommand{\hypersetup}[1]{}
\providecommand{\texorpdfstring}[2]{#1}

\hypersetup{plainpages=false}
\hypersetup{pdfpagemode=UseNone}
\hypersetup{bookmarksnumbered=true}
\hypersetup{pdfstartview=FitH}
\hypersetup{colorlinks=false}
\hypersetup{citebordercolor={.5 1 .5}}
\hypersetup{urlbordercolor={.5 1 1}}
\hypersetup{linkbordercolor={1 .7 .7}}
\providecommand{\arxivref}[2]{\href{http://arxiv.org/abs/#1}{#2}}

\providecommand{\href}[2]{#2}
\providecommand{\arxivlink}[1]{\href{http://arxiv.org/abs/#1}{arxiv:#1}}

\begin{document}

%%%%%%%%%%%%%%%%%%%%%%%%%%%%%%%%%%%%%%%%%%%%%%%%%%%%%%%%%%%%%%%%%%%%%%%%%%%%%%%%
%%%%%%%%%%%%%%%%%%%%%%%%%%%%%%%%%%%%%%%%%%%%%%%%%%%%%%%%%%%%%%%%%%%%%%%%%%%%%%%%
% TITLE PAGE

\thispagestyle{empty}
\phantomsection
\addcontentsline{toc}{section}{Title}

\begin{flushright}\footnotesize%
\texttt{\arxivlink{1012.4000}}\\
overview article: \texttt{\arxivlink{1012.3982}}%
\vspace{1em}%
\end{flushright}

\begingroup\parindent0pt
\begingroup\bfseries\ifarxiv\Large\else\LARGE\fi
\hypersetup{pdftitle={Review of AdS/CFT Integrability, Chapter IV.4: Integrability in QCD and N<4 SYM}}%
Review of AdS/CFT Integrability, Chapter IV.4:\\
Integrability in QCD and $\mathcal{N}<4$ SYM
\par\endgroup
\vspace{1.5em}
\begingroup\ifarxiv\scshape\else\large\fi%
\hypersetup{pdfauthor={G.P. Korchemsky}}%
G.P. Korchemsky
\par\endgroup
\vspace{1em}
\begingroup\itshape
 Institut de Physique Th\'eorique\,\footnote{Unit\'e de Recherche Associ\'ee au CNRS URA 2306},
CEA Saclay, \\
91191 Gif-sur-Yvette Cedex, France
\par\endgroup
\vspace{1em}
\begingroup\ttfamily
Gregory.Korchemsky@cea.fr
\par\endgroup
\vspace{1.0em}
\endgroup

\begin{center}
\includegraphics[width=5cm]{TitleIV4.mps}%figure for your chapter
\vspace{1.0em}
\end{center}

\paragraph{Abstract:}
There is a growing amount of evidence that QCD (and four-dimensional gauge theories in general) possess a hidden symmetry which does not exhibit itself as a symmetry of classical Lagrangians but is only revealed on the quantum level. In this review we consider the scale dependence of local gauge invariant operators and  high-energy (Regge) behavior of scattering amplitudes to explain  that the effective QCD dynamics in both cases is described by completely integrable systems that prove to be related to the celebrated Heisenberg spin chain and its generalizations. 

%\ifarxiv\else
%\paragraph{Mathematics Subject Classification (2010):} 
% http://www.ams.org/msc
%\fi
%\hypersetup{pdfsubject={MSC (2010): ...}}%

\ifarxiv\else
\fi
\hypersetup{pdfkeywords={QCD, anomalous dimensions, scattering amplitudes, integrability}}%

\newpage

%%%%%%%%%%%%%%%%%%%%%%%%%%%%%%%%%%%%%%%%%%%%%%%%%%%%%%%%%%%%%%%%%%%%%%%%%%%%%%%%
%%%%%%%%%%%%%%%%%%%%%%%%%%%%%%%%%%%%%%%%%%%%%%%%%%%%%%%%%%%%%%%%%%%%%%%%%%%%%%%%
% BODY

\newcommand \vev [1] {\langle{#1}\rangle}
\newcommand\re[1]{(\ref{#1})}
\def \qqquad {\qquad\quad}
\def \qqqquad {\qquad\qquad}
\newcommand \Mybf[1] {\mbox{\boldmath$ {#1} $}}
\newcommand \mybf[1] {\mbox{\boldmath$ {\scriptstyle #1} $}}
\newcommand\lr[1]{{\left({#1}\right)}}
\def \be  {\begin{equation}}
\def \ee  {\end{equation}}
\def \ba  {\begin{eqnarray}}
\def \ea  {\end{eqnarray}}
\def \S {\mathop{\rm\bf S}}
\def \Tr {\mathop{\rm Tr}\nolimits}
\def \tr {\mathop{\rm tr}\nolimits}
\newcommand{\bit}[1]{\mbox{\boldmath$#1$}}
\newcommand{\ft}[2]{{\textstyle\frac{#1}{#2}}}
\def \Im {\mathop{\rm Im}\nolimits}
\def \Re {\mathop{\rm Re}\nolimits}

%%%%%%%%%%%%%%%%%%%%%%%%%%%%%%%%%%%%%%%%%%%%%%%%%%%%%%%%%%%%%%%%%%%%%%%%%%%%%%%%
\section{Introduction}

QCD is a four-dimensional gauge theory describing strong interaction of quarks and gluons. There is a growing amount of evidence that QCD (and Yang-Mills theories in general) possess  a hidden symmetry. This symmetry has a dynamical origin in the sense that it is not seen at the level of classical Lagrangian and manifests itself  at quantum level through remarkable integrability properties of effective dynamics. 

\begin{figure}[ht]
%\psfrag{dots}[cc][cc]{$\dots$} 
%\psfrag{sum}[cc][cc]{$\displaystyle \sum_X $} 
%\psfrag{=}[cr][cc]{$\displaystyle F(x,q^2) \ \ =\ \ \Im \hspace*{-3mm} $} 
%\psfrag{2}[cc][cc]{$2$}
%\psfrag{h}[cc][cc]{$h(p)$}  \psfrag{q}[cc][cc]{$\gamma^*(q)$}
%\centerline{{\epsfysize3cm \epsfbox{DIS1.eps}}}
\ifarxiv
\centerline{\includegraphics[height=3cm]{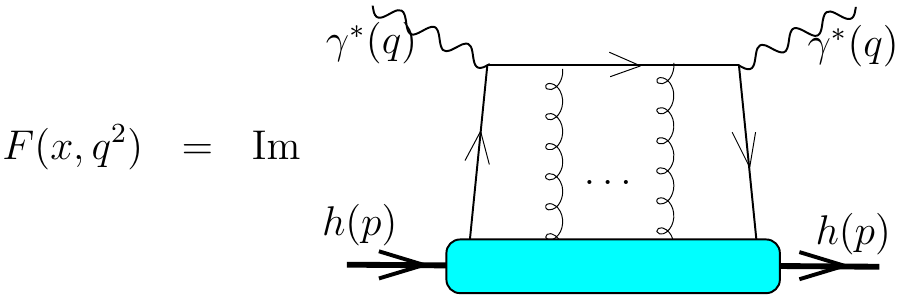}}
\else
\centerline{\includegraphics[height=3cm]{DIS1times}}
\fi
\caption[]{\small The total cross-section of deep inelastic scattering $\gamma^*(q)+h(p)\to {\rm everything}$ is related by the optical theorem to imaginary part of the forward scatteting amplitude. Solid and wavy lines denote quarks and gluons, respectively. }
\label{dis}
\end{figure}

The simplest example which allows us to explain integrability phenomenon is a process of deeply inelastic scattering (DIS) of an energetic hadron off virtual photon, $\gamma^*(q) + h(p) \to \text{everything}$. This process played a distinguished r\^ole in early days of QCD development and it led, in particular, to 
important discoveries such as  QCD factorization and formulation of parton model for hard processes (see e.g. \cite{book}). 
The total cross-section of of DIS process is related by the optical theorem to imaginary part of the forward scatteting amplitude $\gamma^*(q)+h(p)\to\gamma^*(q)+h(p)$  (see Fig.~\ref{dis}). It is parameterized by the so-called structure functions $F(x,q^2)$ depending on the photon virtuality $q^2<0$ and dimensionless Bjorken variable $0<x<1$. The latter is   related to the total center-of-mass energy of the process as $s=(p+q)^2=-q^2(1-x)/x$.

The integrability has been first discovered in Refs.~\cite{Lipatov94,FK95,Korchemsky95} in the study of high-energy, $s\gg -q^2$ (or equivalently $x\to 0$) asymptotics of $F(x,q^2)$.
Experimental data indicate that the structure functions  increase in this limit as a power of the energy, $F(x,q^2)\sim (1/x)^{\omega}$, in a quantitative agreement with the Regge theory prediction. At weak coupling, the same behavior can be obtained through resummation of perturbative corrections to the structure functions enhanced by logarithm of the energy~\cite{BFKL}. The structure functions obtained in this way satisfy nontrivial multi-particle Bethe-Salpeter like evolution equations~\cite{Bartels80,KP80}.
These equations have resisted analytical solution but a breakthrough occurred after it was found \cite{Lipatov94,FK95,Korchemsky95} that, in multi-color limit, these equations can be mapped into a Sch\"odinger equation for a completely integrable quantum (noncompact) Heisenberg  $SL (2, \mathbb{C})$ spin chain. This opened up the possibility of applying the quantum inverse scattering methods for the construction of the exact solution to the evolution equation in planar QCD. 

Later, similar integrable structures have been found in Refs.~\cite{BDM98,Belitsky99,BDKM99} in the study of dependence of the structure functions $F(x,q^2)$ on the momentum transfered $q^2$. At large $Q^2=-q^2$, the operator product expansion can be applied to expand the moments of the structure functions
in powers of a hard scale $1/Q$
\begin{align}\label{OPE}
\int_0^1 dx\, x^{N-1} F(x,q^2) =\sum_{L\ge 2} \frac{c_{N,L}(\alpha_s(Q^2))}{Q^L} \vev{p|O_{N,L}|p}_{\mu^2=Q^2} \,.
\end{align}
Here the expansion runs over local composite gauge invariant operators ({\it Wilson operators}) of  Lorentz spin $N$ and twist $L$. The corresponding coefficient functions $c_{N,L}(\alpha_s(Q^2))$ can be computed at 
weak coupling as a series in the QCD coupling constant $\alpha_s(\mu^2)=g^2/(4\pi)$ normalized at 
$\mu^2=Q^2$. At the same time, the matrix element of the Wilson operator with respect to
hadron state $\vev{p|O_{N,L}|p}_{\mu^2=Q^2}$ is a nonperturbative quantity. Its absolute value can not be computed perturbatively whereas its dependence on the hard scale $Q^2$ is governed by 
the renormalization group (Callan-Symanzik) equations
\begin{align}\label{CS}
\mu^2\frac{d}{d\mu^2}   \vev{p|O^{(\alpha)}_{N,L}|p} = - \gamma^{(\alpha)}_{N,L}(\alpha_s) \vev{p|O^{(\alpha)}_{N,L}|p} \,.
\end{align}
Here we introduced the superscript $(\alpha)$ to indicate that for given $N$ and $L$ there are a few Wilson operators parameterized by the index $\alpha$. The Callan-Symanzik equation \re{CS} has the meaning of  a conformal Ward identity for the Wilson operators with the anomalous dimension $ \gamma^{(\alpha)}_{N,L}(\alpha_s) $ being the eigenvalue of the QCD dilatation operator (see e.g. review \cite{BKM03}).  
 
The Wilson operators are built in QCD from elementary quark and gluon fields and from an arbitrary number of covariant derivatives. In general, such operators mix under renormalization with other operators carrying the same Lorentz spin and twist. Diagonalizing
the corresponding mixing matrix we can find the spectrum of the anomalous dimensions $\gamma^{(\alpha)}_{N,L}(\alpha_s)$. For the Wilson operators of the lowest twist, $L=2$, the anomalous dimensions can be obtained in the closed form \cite{GW73}, whereas for higher twist operators the problem becomes extremely nontrivial already at one loop
due to a complicated form of the mixing matrix \cite{BFLK85}. Quite remarkably, the spectrum of 
the anomalous dimensions can be found exactly in QCD in the sector of the so-called maximal-helicity Wilson operators. The reason for this is that the one-loop mixing matrix in QCD in this sector can be mapped in the multi-color limit into a Hamiltonian of the Heisenberg $SL(2 ,\mathbb{R})$ spin chain~\cite{BDM98,Belitsky99,BDKM99}. The twist of the Wilson operator $L$ determines the length of the spin chain while the spin operators in the each site are defined by the generators of the `collinear' $SL(2 ,\mathbb{R})$ subgroup of the full conformal group~\cite{Makeenko:bh,Ohrndorf82}. As a result,   the exact spectrum of one-loop anomalous dimensions can be computed with a help of Bethe Ansatz~\cite{QISM}. 

Let us now examine the relation \re{OPE} for large Lorentz spin, $N\gg 1$. This limit has important phenomenological applications in QCD~\cite{Simula:2000ka,Gardi:2002xm}. It is known \cite{GW73} that the anomalous dimensions of Wilson operators grow as their Lorentz spin increases. As a consequence, the dominant contribution to \re{OPE} only comes from the operators with the minimal anomalous dimension $\gamma^{(0}_{N,L} = \min_{\alpha}  \gamma^{(\alpha)}_{N,L}$.
Quite remarkably, this anomalous dimension has a universal (twist $L$ independent) logarithmic scaling behavior at large $N$ to all loops~\cite{Korchemsky88,bgk}
\begin{align}\label{min}
 \gamma^{(0)}_{N,L} =  2 \Gamma_{\rm cusp}(\alpha_s) \ln N + O(N^0)\,,
\end{align}
where  $\Gamma_{\rm cusp}(\alpha_s)$ is the \textit{cusp anomalous dimension} \cite{Korchemsky:1987wg}. 

By definition, $\Gamma_{\rm cusp}(\alpha_s)$ governs the scale dependence
of Wilson lines with light-like cusps \cite{polyakov2,Korchemskaya:1992je} and its relation to anomalous dimensions of large
spin Wilson operators is by no means obvious. 
It can be understood \cite{Korchemsky88} by invoking the physical picture of deep inelastic scattering at large $N$. In terms of the moments \re{OPE}, large $N$ corresponds to the region of $x\to 1$.
For $x\to 1$ the final state in the deep inelastic scattering has a small invariant mass, $s=Q^2(1-x)/x \ll Q^2$, and it consists of a collimated jet of energetic particles accompanying by soft gluon radiation.
Interacting with soft gluons, the particles inside the jet acquire the eikonal phases given by
Wilson line operators $P\exp(i\int_0^\infty dt\, p\cdot A(pt))$ evaluated along semi-infinite line  in the direction of the particle momenta. In this way, for $x\to 1$,  complicated QCD
dynamics in deep inelastic scattering  admits an effective description in terms 
of Wilson lines~\cite{Ivanov:1985np}. The relation \re{min} between anomalous dimensions and cusp singularities
of light-like Wilson lines is just one of the application of this formalism. Another examples
include the relation between light-like Wilson loops with  on-shell scattering amplitudes,
Sudakov form factors, gluon Regge trajectories etc (see Ref.~\cite{Drummond:2007aua} and references
therein).
  
At present, integrability of the dilatation operator in planar QCD has been verified to two loops in the $SL(2;\mathbb{R})$ sector of maximal helicity operators~\cite{Belitsky:2006av}. In other sectors, the dilatation operator receives additional contribution that breaks integrability  already to one loop. This contribution vanishes however for large values of the Lorentz spin $N\gg 1$ thus suggesting that integrability in planar QCD gets restored to all loops in the leading large $N$ limit~\cite{GKK02}. Indeed, as was shown in Ref.~\cite{bgk}, the all-loop dilatation operator in QCD in the $SL(2;\mathbb{R})$ sector can be mapped in the large $N$ limit into a Hamiltonian of a {\em classical}  Heisenberg $SL(2;\mathbb{R})$ spin chain. In this manner, the Wilson operators
with large $N$ are described by the so-called finite-gap solutions and the spectrum of anomalous dimension can be found through their semiclassical quantization. In 
particular, the relation \re{min} naturally appears as describing the ground state energy
of the classical $SL(2;\mathbb{R})$ spin chain of an arbitrary length $L$ and total
spin $N$.

The above mentioned integrability structures (those of the scattering amplitudes in the Regge limit
and of the dilatation operator) are not specific to QCD. They are
also present in generic four-dimensional gauge theories including supersymmetric Yang-Mills
models with $\mathcal{N}=1,2,4$ supercharges. Supersymmetry enhances the
phenomenon by extending integrability to a larger class of observables.
In this context, the maximally supersymmetric $\mathcal{N}=4$ Yang-Mills
theory is of a special interest with regards to the AdS/CFT correspondence
\cite{maldacena}.   The gauge/string
duality hints that these structures should manifest themselves through hidden
symmetries of the scattering amplitudes and of anomalous dimensions in dual
gauge theories to all loops.

\section{Integrability of dilatation operator in QCD}

In this section, we review a hidden integrability of the dilatation operator in a generic
four-dimensional  Yang-Mills theory describing the coupling of gauge fields
to fermi\-ons and scalars. Depending on the representation in which the latter
fields are defined, we can distinguish two different types of the gauge
theories: QCD and supersymmetric extensions of Yang-Mills theory (SYM). 

In QCD, the gauge fields are coupled to quarks in the fundamental
representation of the $SU(N_c)$ gauge group.
The quarks are described by four-component Dirac fermions $\psi$ and the gauge field strength
$F_{\mu\nu}=\frac{i}{g}[D_\mu,D_\nu]$ is determined in terms of the covariant
derivatives $D_\mu = \partial_\mu - i g A_\mu^a t^a$ with generators
$t^a$ in the fundamental representation of the $SU(N_c)$ normalized
conventionally as ${\rm tr} \, (t^a t^b) = \frac12 \delta^{ab}$.
In SYM theory, the gauge fields are coupled to fermions (gauginos) and scalars belonging to the adjoint representation of the $SU(N_c)$ group.
The supersymmetric Yang-Mills theories with $\mathcal{N}=1,2$ and $4$ supercharges are obtained from the  Lagrangian of generic Yang-Mills theory by adjusting the number of gaugino and scalar species.
The gauginos are described by the Weyl fermion $\lambda^A$ which belongs to
the fundamental representation of an internal $SU(\mathcal{N})$ symmetry group
with its complex conjugate $\bar \lambda_A=(\lambda^A)^*$. The scalars are
assembled into the antisymmetric tensor $\phi^{AB} = - \phi^{BA}$, with its
complex conjugate $( \phi^{AB} )^\ast = \bar\phi_{AB}$.  As we explain below,
integrability is not tied to supersymmetry and the phenomenon persists in the 
generic Yang-Mills theory  for arbitrary  $\mathcal{N}$, to two loop order at least.
 
\subsection{Light-ray operators} 
 
Let us first consider renormalization of local gauge invariant operators in QCD.  As the simplest example, we examine the following twist-two operator contributing to the moments of DIS structure function \re{OPE}
\begin{align}\label{qq}
\vev{p|O_{N,L=2}(0)|p} = \vev{p|\bar\psi\, \gamma_+ D_+^{N-1} \psi(0)|p}\,.
\end{align}
It is built from two quark fields and $(N-1)$ covariant derivatives $D_+=(n\cdot D)$ projected onto light-like vector $n_\mu=q_\mu-p_\mu  q^2/(2 pq)$ and $\gamma_+=(n\cdot \gamma)$ being the projected Dirac matrix. Discussing renormalization properties
of  Wilson operators like \re{qq} it is convenient to switch from infinite set
of local operators \re{qq} parameterized by positive integer $N$ to a single
nonlocal {\it light-ray operator}
\begin{align}\label{O-def}
\mathbb{O}(z_1,z_2) = \bar \psi(z_1 n) \gamma_+ [n z_1, n z_2] \psi(z_2 n)
= \sum_{N\ge 1}\left[ \bar\psi\, \gamma_+ D_+^{N-1} \psi\right] \frac{(z_1-z_2)^{N-1}}{(N-1)!}+\ldots
\end{align}
Here $z_1$ and $z_2$ are scalar variables defining the position of quark fields
on the light-cone and the gauge link $[n z_1, n z_2]\equiv P\exp(ig\int_{z_1}^{z_2} dt \,A_+(nt))$ is inserted to ensure gauge invariance of $\mathbb{O}(z_1,z_2)$. Also,
ellipses in the right-hand side of \re{O-def} stand for terms involving total derivatives
of the twist-two operators and, therefore, providing vanishing contribution to 
the forward matrix element $\vev{p|\mathbb{O}(z_1,z_2)|p}$.

We recall that local gauge invariant operators
% with the same quantum numbers mix under renormalization and 
satisfy the evolution equation \re{CS}. The same is true for the
light-ray operators \re{O-def} although the explicit form of the evolution equation is different due to nonlocal form of the light-ray operators. In particular, for the operators
\re{O-def} the evolution equation takes the following form 
\cite{AZ78,BFLK85,BB89,MRGDH94}
\be
\left(
\mu \frac{\partial}{\partial \mu} + \beta(g^2) \frac{\partial}{\partial g^2}
\right) \mathbb{O} (z_1, z_2)
=
- [ \mathbb{H}_2(g^2) \cdot \mathbb{O}] (z_1, z_2)
\, ,
\label{RG}
\ee
with  the evolution kernel $\mathbb{H}_2$ to be specified below.
The evolution equation \re{RG} expresses the conformal Ward identity in QCD and the beta-function term takes into account conformal symmetry breaking contribution. 
The evolution operator $\mathbb{H}_2$ in the right-hand side of \re{RG} defines a representation of the dilatation operator on the space spanned by nonlocal light-ray operators \re{O-def}.  In general,
$\mathbb{H}_2$ has a matrix form as the light-ray operators with different
partonic content could mix with each other.

The evolution kernel $\mathbb{H}_2$ has a perturbative expansion in powers of the coupling constant and
admits a representation in the form of an integral operator acting on light-cone coordinates $z_1$ and $z_2$ of  $\mathbb{O}(z_1, z_2)$.
To the lowest order in the coupling, its explicit form   
has been found in QCD in Ref.~\cite{BB89} and its generalization to
Yang-Mills theories with an arbitrary number of supercharges has been derived in Ref.~\cite{BDKM04}. The corresponding expressions for $\mathbb{H}_2$ are given below in Eq.~\re{H-tw-2} .
The main advantage of \re{RG} compared with the conventional approach based on
explicit diagonalization of the mixing matrix for local Wilson operators is that the
problem of finding the spectrum of anomalous dimensions can be mapped into
spectral problem for one-dimensional quantum mechanical Hamiltonian $\mathbb{H}_2$. As we will see in a moment, the same happens in QCD
for Wilson operators of high twist $L\ge 3$, in which case the corresponding evolution operator $\mathbb{H}_L$ in the sector of maximal helicity operators
turns out to be equivalent for a Hamiltonian of Heisenberg $SL(2;\mathbb{R})$ spin chain of length $L$.

\subsection{Light-cone formalism}

Discussing integrability of the dilatation operator in QCD and in  SYM theories, it 
is convenient to employ the ``light-cone formalism'' \cite{KogSop,BLN83,Mandelstam83}. In this
formalism one integrates out non-propagating components of fields and formulates
the (super) Yang-Mills action in terms of ``physical'' degrees of freedom.
Although the resulting action is not manifestly covariant under the Poincar\'e
transformations, the main advantage of the light-cone formalism for SYM theories
is that the $\mathcal{N}-$extended supersymmetric algebra is closed off-shell for
the propagating fields and there is no need to introduce auxiliary fields. This
allows us to design a unifying light-cone superspace formulation of various
$\mathcal{N}-$extended SYM, including the ${\cal N} = 4$ theory for which a
covariant superspace formulation does not exist.

In the light-cone formalism, one quantizes the Yang-Mills theory in a noncovariant,
light-cone gauge $(n\cdot A) \equiv A_+(x) = 0$. Introducing an auxiliary
complimentary light-like vector $\bar n_\mu$, 
such that $\bar n^2=0$ and $(n\cdot \bar n)=1$, we split three remaining components of the gauge field  into longitudinal, $A_- (x)$, and two
transverse holomorphic and antiholomorphic components, $A(x)$ and $\bar A(x)$,
respectively,
\begin{equation}
A_- \equiv (\bar n \cdot A)\,,\qquad A \equiv \ft1{\sqrt{2}}(A_1 +
i A_2) \, , \qquad
\bar A \equiv A^* = \ft1{\sqrt{2}}(A_1 - i A_2)
\, .
\label{gauge-perp}
\end{equation}
In the similar manner, the 
fermion field ${\psi}(x)$ can be decomposed 
%into the so-called ``bad'' and ``good'' components 
with a help of projectors $ \Pi_\pm = \ft12 \gamma_\pm
\gamma_\mp$  as % (with $\Pi_\pm^2=\Pi_\pm$ and $\Pi_\pm\Pi_\mp=0$)
\be
{\psi} = {\Pi}_+ {\psi} + {\Pi}_- {\psi} \equiv {\psi}_+ + {\psi}_-\,,
\label{good-bad}
\ee
where the fermion field ${\psi}_+$ has two nonzero components
\be
q_{ \uparrow}
= \frac12 ( 1 - \gamma_5 ) \psi_+
\, , \qquad q_{\downarrow} = \frac12 ( 1 + \gamma_5 )  
\psi_+ \,.
\label{1/2-QCD}
\ee
Then, one finds that the fields ${\psi}_-(x)$ and $A_-(x)$ can be integrated out
and the resulting action of the Yang-Mills theory is expressed in terms of
``physical'' fields: complex gauge field, $A(x)$ and $\bar A(x)$, two components of fermion fields, $q_{ \uparrow}(x)$ and $q_{ \downarrow}(x)$, and, in the case of supersymmetric gauge theory, complex scalar fields $\phi(x)$. When applied to the vacuum states, the fields  $(A,\,q_{ \downarrow},\, \phi,\, q_{ \uparrow},\,\bar A)$
create massless particles of helicity $(-1,-\ft12,0,\ft12,1)$, respectively. 

Taking the product of `physical' fields and light-cone derivatives $D_+=\partial_+$, 
we can construct the set of local gauge invariant operators.  Such operators define
the representation of the so-called collinear $SL(2;\mathbb{R})$ subgroup of the conformal group and they are known in QCD literature as  \textit{quasipartonic operators}. 
A distinguished feature of these operators is that their twist equals the number of constituent physical fields~\cite{BFLK85}. In analogy with  \re{O-def}, we can 
replace an infinite number of Wilson operators of a given twist $L$ with a few 
nonlocal light-ray operators $\mathbb{O}(z_1,\dots , z_L)$. The latter can be thought of
as generating functions for the former. Due to different
$SU(N_c)$ representation of fermions (fundamental in QCD and adjoint in SYM), the definition of such operators
is slightly different in the two theories. 

In QCD, in the simplest
case of twist two, we can distinguish four different light-ray operators
(plus complex conjugated operators)
\begin{align}\notag
& \mathbb{O}_{qq}^{(0)}(z_1, z_2) = \bar q_{\uparrow}(n z_1) q_{ \uparrow}(n z_2) \,,\quad
\mathbb{O}_{gg}^{(0)}(z_1, z_2) = \tr\left[ \partial_+ \bar A(n z_1) \partial_+ A(n z_2)\right]\,, 
\\[2mm]
& \label{tw-2}
\mathbb{O}_{qq}^{(1)}(z_1, z_2) = \bar q_{\downarrow}(n z_1) q_{ \uparrow}(n z_2) \,, \quad
 \mathbb{O}_{gg}^{(2)}(z_1, z_2) = \tr\left[ \partial_+ A(n z_1) \partial_+ A(n z_2)\right] \,,
\end{align}
where the subscript ($qq$ and $gg$) indicates particle content of the operator
and the superscript defines the total helicity. In this basis, the operator
\re{O-def} is given by a linear combination of $ \mathbb{O}_{qq}^{(0)}(z_1, z_2)$ and complex conjugated operator.
The operators $\mathbb{O}_{qq}^{(0)}$ and $\mathbb{O}_{gg}^{(0)}$ have the same quantum numbers and mix under
renormalization. At the same time, the operators $\mathbb{O}_{qq}^{(1)}$ and $\mathbb{O}_{gg}^{(2)}$ carry different helicity and have an autonomous
scale dependence. In what follows we shall refer to them as maximal helicity
operators. The reason why we distinguish such operators is that the one-loop dilatation operator in QCD is integrable in the sector of maximal helicity
operators only.

For higher twist $L\ge 3$ we can define three different types of maximal
helicity operators in QCD:
\begin{align}
&  \mathbb{O}_{qqq}^{(3/2)} (z_1, z_2, z_3)
 =
\varepsilon_{ijk} \, q_{\uparrow}^i (z_1 n) q_{\uparrow}^j (z_2 n) q_{\uparrow}^k (z_3 n)
\, , \label{O3-QCD}
\\\label{O-mixed}
&  \mathbb{O}_{qg \ldots g q}^{(L-1)} (z_1, \ldots, z_L)
= \bar q_{\downarrow}(n z_1) \partial_+ A(nz_2)\ldots \partial_+ A(nz_{L-1})
q_{ \uparrow}(n z_L) \, ,
\\\label{ggg}
& \mathbb{O}_{g \ldots g }^{(L)} (z_1, \ldots, z_L)
= \tr\left[ \partial_+ A(nz_1)\ldots \partial_+ A(nz_{L})\right]\, ,
\end{align}
to which we shall refer as baryonic  $(L=3)$ operators, mixed quark-gluon operators and
gluon operators, respectively.
We remind that since quark fields belong to the fundamental representation of the $SU(N_c)$ group, the length of the operator \re{O3-QCD} ought to be $N_c=3$.  At the same time, gluon fields are in the adjoint representation and the single trace
operator \re{ggg} is well-defined for arbitrary $N_c$ and twist $L$. The same applies to the mixed quark-gluon operators \re{O-mixed}.
The operators \re{O3-QCD} and \re{O-mixed} have a direct phenomenological significance: their matrix elements determine the
distribution amplitude of the delta-isobar~\cite{BroLep79} and higher twist contribution to spin structure functions, respectively.

\subsection{Evolution kernels}

The light-ray operators \re{tw-2} -- \re{ggg} satisfy the evolution equation \re{RG}. Let us first examine  twist-two quark operators $\mathbb{O}_{qq}^{(0)}$ and $\mathbb{O}_{qq}^{(1)}$  defined in \re{tw-2}. The operator $\mathbb{O}_{qq}^{(0)}$ can mix with the gluon operator $\mathbb{O}_{gg}^{(0)}$. To simplify the situation,
we can suppress the mixing by choosing the two quark fields inside $\mathbb{O}_{qq}^{(0)}$ to have different flavor. To one-loop order, the evolution kernel receives
the contribution from one-gluon exchange between two quark fields and from self-energy corrections. The latter one is the same for the two operators while the former one is different
\begin{align} \notag
\mathbb{H}^{(1)}_{qq} &=\frac{g^2 C_F}{8\pi^2}\left[H_{12} + 2 \gamma_q \right]\,,
\\[2mm] \label{H-tw-2}
\mathbb{H}^{(0)}_{qq}&=\frac{g^2 C_F}{8\pi^2}\left[H_{12} + V_{12} + 2 \gamma_q \right]\,.
\end{align}
Here $C_F=t^at^a=(N_c^2-1)/(2N_c)$ is the quadratic Casimir of the $SU(N_c)$ in the fundamental representation, $\gamma_q=1$ is one-loop anomalous dimension
of quark field in the axial gauge $A_+=0$ and $H_{12}$ and $V_{12}$ are integral operators
\begin{align}\notag
[  {H}_{12}  \cdot
\mathbb{O} ] (z_1, z_2) &= \int_0^1 \frac{d \alpha}{\alpha} \bar\alpha
%^{2 j -1} {}
\Big[ 2 \mathbb{O} (z_1, z_2) - \mathbb{O} (\bar\alpha z_1 + \alpha
z_2, z_2) - \mathbb{O} (z_1, \alpha z_1 + \bar\alpha z_2) {}\Big],
\\\label{LightConeKernel}
[  {V}_{12}  \cdot
\mathbb{O} ] (z_1, z_2) &= \int_0^1 d\alpha_1\int_0^{\bar\alpha_1}d\alpha_2\,
\mathbb{O}(\alpha_1 z_1 + \bar\alpha_1 z_2,\alpha_2 z_2 + \bar\alpha_2 z_1)\,,
\end{align}
where $\bar\alpha_i \equiv 1 - \alpha_i$. These operators have a transparent
physical interpretation: they displace two particles along the light-cone in the direction of each other.

To find the spectrum of anomalous dimensions of twist-two quark operators
generated by light-ray operators \re{tw-2}, we have to diagonalize 
the operators $\mathbb{H}^{(1)}_{qq}$ and $\mathbb{H}^{(0)}_{qq}$. This can be 
done with a help of conformal symmetry. We recall that the conformal symmetry
is broken in QCD at loop level. However the dilatation operator receives 
conformal symmetry breaking contribution only starting from two loops and, as a consequence, the one-loop evolution kernels in QCD have to respect  conformal symmetry of QCD Lagrangian. For nonlocal light-ray operators built from fields $X(n z)$,
the full $SO(2,4)$ conformal symmetry reduces to its collinear $SL(2;\mathbb{R})$
subgroup acting on one-dimensional light-cone coordinates of fields~\cite{Makeenko:bh,Ohrndorf82}
\begin{align}
z \to \frac{az+b}{cz+d}\,,\qquad X(z n ) \to (cz+d)^{-2j} X\left(\frac{az+b}{cz+d}n  \right)
\end{align}
with $ad-bc=1$. The generators of these transformations are 
\begin{align}\label{l-col}
L^-=-\partial_z\,,\qquad L^+=2j z+ z^2\partial_z\,,\qquad L^0=j+z\partial_z\,.
\end{align}
Here $j$ is the conformal weight of the field. For  `physical'  components of fermions, $\psi_+$, it equals $j_q=1$, for transverse components
of gauge field, $\partial_+ A$ and $\partial_+\bar A$, it is $j_g=3/2$ and for scalars 
$j_s=1/2$.

In application to light-ray quark operators, $\mathbb{O}_{qq}^{(0)}(z_1, z_2)$ and $\mathbb{O}_{qq}^{(1)}(z_1, z_2)$, the conformal symmetry dictates that the one-loop evolution kernels \re{H-tw-2} have to commute with the two particle conformal spin 
$ L_1^\alpha+L_2^\alpha$ (with $\alpha=-,+,0$).
As a consequence, $\mathbb{H}_{qq}^{(h=0,1)}$ is a function of the corresponding two-particle Casimir operator
\begin{align}
L_{12}^2=\sum_{\alpha=+,-,0} (L_1^\alpha+L_2^\alpha)^2 = J_{12}(J_{12}-1)\,.
\end{align}
To find the explicit form of this dependence, it suffices to examine the action of the two
operators, $\mathbb{H}_{qq}^{(h)}$ and $L_{12}^2$, on the same test function $(z_1-z_2)^{n}$, which is just the lowest weight in the tensor product of two $SL(2;\mathbb{R})$ representations carrying the spin $J_{12}=n+2$. Replacing $\mathbb{O}(z_1,z_2)\to (z_1-z_2)^{J_{12}-2}$ in \re{LightConeKernel} we
find
\begin{align}\label{h12}
H_{12} = 2\left[\psi(J_{12})-\psi(2)\right]\,,\qquad V_{12} = 1/(J_{12}(J_{12}-1))\,,
\end{align}
where $\psi(x)=d\ln\Gamma(x)/dx$ is Euler psi-function. Together with \re{H-tw-2} these relations determine the spectrum of anomalous dimensions of twist-two quark operators.

\subsection{Relation to Heisenberg \texorpdfstring{$SL(2;\mathbb{R})$}{SL(2;R)} spin chain}

As the first sign of integrability, we notice that  $H_{12}$ coincides
with the known expression for two-particle Hamiltonian of Heisenberg spin chain~\cite{KRS81,TTF83}
\begin{align}\label{closed}
H_L=H_{12}+\ldots+H_{L1}\,,\qquad H_{i,i+1} = \psi(J_{i,i+1})-\psi(2j)\,,
\end{align}
where the spin operators are identified as $SL(2;\mathbb{R})$  conformal generators \re{l-col}. As follows from \re{H-tw-2}, the one-loop dilatation operator $\mathbb{H}_{qq}^{(1)}$
depends on $H_{12}$ and, therefore, it is mapped into Heisenberg $SL(2;\mathbb{R})$ spin chain of length 2. At the same time, the dilatation operator $\mathbb{H}_{qq}^{(0)}$ receives the additional contribution $V_{12}$. It preserves the conformal symmetry but breaks
integrability. Notice that $V_{12}$ vanishes for large values of the conformal spin $J_{12}\gg 1$ so that the two evolution kernels, $\mathbb{H}_{qq}^{(0)}$ and $\mathbb{H}_{qq}^{(1)}$, have the same asymptotic behavior at 
large $J_{12}$.  This suggests that for the operator $\mathbb{H}_{qq}^{(0)}$ integrability is restored in the limit of large conformal spin only. 

For twist-two operators, the anomalous dimensions are uniquely determined by their conformal spin. To appreciate the power of integrabilty, we have to consider
Wilson operators of high twist $L\ge 3$. For example, for the maximal helicity baryonic operators \re{O3-QCD} the one-loop dilatation operator has the form \cite{BDKM99}
\begin{align}
\mathbb{H}_{qqq}^{(3/2)} = \frac{\alpha_s}{2\pi}\left[ (1+1/N_c) (H_{12}+H_{23}+H_{31})+\frac32 C_F\right]
\end{align}
with $N_c=3$ and $H_{12}$ given by \re{h12}. Comparing this relation with \re{closed} we recognize that $\mathbb{H}_{qqq}^{(3/2)}$ can be mapped into
Heisenberg spin chain of length $L=3$. The spin at each site $j=1$ is determined
by the conformal spin of quark field. 

For gluon operators of the maximal helicity \re{ggg} the dilatation operator receives
contribution from self-energy corrections to gluon fields and from one-gluon exchange between any pair of gluons. The latter produces both planar and nonplanar corrections (for $L>3$). In the planar limit, the one-loop dilatation operator has the following form~\cite{Belitsky00}
\begin{align}
\mathbb{H}_{g \ldots g }^{(L)} = \frac{g^2 N_c}{8\pi^2} (H_{12}+\ldots+H_{L1})\,,
\end{align}
where two-particle kernel $H_{i,i+1}$ acts locally on light-cone coordinates of
gluons with indices $i$ and $i+1$. The conformal symmetry implies that $H_{i,i+1}$
is a function of the conformal spin of two gluons $J_{i,i+1}$. Quite remarkably, the
dependence of $H_{i,i+1}$ on $J_{i,i+1}$ has the same form as in \re{closed}. As a consequence, the one-loop planar dilatation operator
for maximal helicity gluon operator \re{ggg} coincides with the Hamiltonian of the 
Heisenberg $SL(2;\mathbb{R})$ spin chain. The length of the spin chain equals
the twist of the operator $L$ and the spin in each site $j=3/2$ coincides with the 
conformal spin of the gluon field.

For mixed quark-gluon operators of the maximal helicity \re{O-mixed}, the quark fields can 
interact in the planar limit with the adjacent gluon fields only  while quark-quark
interaction is suppressed in this limit. As a consequence, the one-loop dilatation
operator has the following form in the planar limit
\begin{align} \label{open}
\mathbb{H}_{qg \ldots gq}^{(L-1)} = \frac{g^2 N_c}{8\pi^2} (U_{12} + H_{23}+\ldots+H_{L-1,L}+U_{L-1,L})\,.
\end{align}
Here $H_{i,i+1}$ describes the interaction of two gluons with aligned helicities and
it is the same as in \re{closed}. The kernels $U_{12}$ and $U_{L-1,L}$ describes quark-gluon interaction
 and their explicit form can be found in Ref.~\cite{DKM00,Belitsky:1999bf}. Notice that the operator $\mathbb{H}_{qg \ldots gq}^{(L-1)}$ has the form of a Hamiltonian of {\it open spin chain} of length
$L$. The spin in sites $1$ and $L$ coincides with the conformal spin of
quark $j_q=1$ and the spin in all remaining sites is given by gluon conformal spin 
$j_g=3/2$. As was shown in Ref.~\cite{DKM00,Belitsky:1999bf}, the open spin chain \re{open} is integrable.
 
\subsection{Exact solution}

Integrability of the one-loop dilatation operator allows us to find
the exact spectrum of anomalous dimensions with a help of the Bethe Ansatz
\cite{BDM98,Belitsky99,BDKM99}
\begin{align} \notag
& \gamma_{N,L} = \frac{g^2 N_c}{8\pi^2} {\mathcal E}_{N,L} + O(g^4)\,,
\\
%\end{align}
%In the case of the closed spin chain Hamiltonian \re{closed}, the exact
%expression for the energy can be written in two equivalent forms
%\begin{equation}
& {\mathcal E}_{N,L}
= \sum_{k=1}^N \frac{2j}{u_k^2 + j^2}= i \frac{d}{d u}\ln \frac{Q(u + ij)}{Q(u - ij)}\bigg|_{u=0}\, .
\label{E-Q}
\end{align}
Here $j$ is the conformal spin in each site ($j=1$ for quark operators and $j=3/2$ for gluon operators),
$u_k$ are Bethe roots and $Q(u)$ is a polynomial of degree $N$ of the form
\begin{equation}
Q(u)=\prod_{j=1}^N (u-u_j)\, .
\label{Bethe}
\end{equation}
The function $Q(u)$ defined in this way has the meaning of the eigenvalue of the Baxter operator for the
$SL(2,\mathbb{R})$ magnet~\cite{Korchemsky95,Derkachov99}. It satisfies the finite-difference
Baxter equation
\begin{equation}
t_L(u) Q(u) = (u+ij)^L Q(u+i) + (u-ij)^L Q(u-i)\,,
\label{Baxter-eq}
\end{equation}
where $t_L(u)$ is the transfer matrix of the spin chain
\begin{equation}
t_L(u) = 2u^L+q_2 u^{L-2} +\ldots+ q_L
\label{transfer}
\end{equation}
and $q_2,\ldots,q_L$ are the conserved charges.

The Baxter equation \re{Baxter-eq} alone does not specify $Q(u)$ uniquely and it has to be supplemented
by additional condition for analytical properties of $Q(u)$. For the $SL(2;\mathbb{R})$ spin
chains describing the anomalous dimensions, $Q(u)$ has to be a polynomial in the spectral parameter.  
%\footnote{This is equivalent to the requirement for nonlocal light-ray operator \re{O-def} to have a series expansion in powers of $z$'s.}
Being combined with the Baxter equation \re{Baxter-eq},
this condition determines $Q(u)$ up to an overall normalization and, as a consequence, allows us to establish the quantization conditions for the $q-$charges and to compute the
exact energy ${\mathcal E}_{N,L}$.

Solving the Baxter equation \re{Baxter-eq} for $N=0,1,\ldots$ one finds the eigenspectrum of the Hamiltonian $\mathbb{H}_{L}$ and, as
a consequence, determines the exact spectrum of the anomalous dimensions
of the maximal helicity baryon operators (for $j=1$ and $L=3$) and of
maximal helicity gluon operators (for $j=3/2$ and $L\ge 2$). 
The spectrum obtained in this way exhibits remarkable regularity: almost all eigenvalues are double degenerate and for large $N$ they
belong to the set of trajectories~\cite{Korchemsky95,Korchemsky:1995be}. Both properties are ultimately related to integrability
of the dilatation operators and can be served to test integrability at high loops. 

For the $SL(2;\mathbb{R})$ spin chains under consideration, the Baxter equation approach and conventional Bethe Ansatz are equivalent. Indeed, substituting \re{Bethe} into the Baxter equation \re{Baxter-eq}, one finds that the roots $u_j$ satisfy the conventional $SU(2)$ Bethe equations for spin $(-j)$. The fact that the spin is negative leads to a number of important differences as compared to   ``compact''
$SU(2)$ magnets. In particular, the Bethe roots take real values only and the
number of solutions is infinite~\cite{Korchemsky95,Korchemsky:1995be}.

\subsection{Semiclassical limit}

  The Baxter operator approach
becomes advantageous when one studies the properties of anomalous dimensions
at large spin $N$ and/or twist $L$. The reason for this is that the Baxter equation
\re{Baxter-eq} takes the form of discretized Schr\"odinger equation. After rescaling
of the spectral parameter, $u\to (N+Lj) x$, we can seek for solution to  \re{Baxter-eq} in the WKB form~\cite{PG92,Sklyanin,Korchemsky:1995be} 
\begin{align}\label{WKB}
Q(Nx) = \exp\left(\frac{i}{\hbar} S(x)\right)\,,\qquad \hbar =1/(N+Lj)\,,
\end{align}
where the action function $S(x)$ admits an expansion in powers of  $\hbar$. Substitution of \re{WKB} into the Baxter equation \re{Baxter-eq} yields the equation for $S(x)$ which can be solved as a series in $\hbar$. To leading order we have
\begin{align}
S(x) = \int_{x_0}^x dx\, p(x) + O(\hbar)\,,
\end{align}
where the momentum $p(x)$ is defined on the spectral curve (``equal energy'' condition) of the classical $SL(2;\mathbb{R})$ magnet $y(x)=2x^L\sinh p(x)$ with
\cite{Korchemsky:1996kh}
\begin{equation}
\Gamma_L: \qquad y^2=(t_L(x))^2 - 4x^{2L}\,.
\label{curve}
\end{equation}
The classical dynamics on this spectral curve has been studied in detail in Refs.~\cite{Korchemsky:1997yy,GKK02}.
Using \re{WKB} we can compute the asymptotic behavior of the energy as~\cite{Korchemsky:1995be,Belitsky:2006en}
 \begin{align}\label{E-as}
\mathcal{E}_{N,L}^{\rm (as)} = 2\ln 2 + \sum_{n=1}^L \left[ \psi(j+i\delta_n)+ \psi(j-i\delta_n)
-\psi(2j)\right] +\ldots\,,
\end{align}
where ellipses denote terms subleading at large $(N+jL)$. Here  $\delta_n$ are roots
of the transfer matrix defined in \re{transfer},  $t_L(\delta_n)=0$. They depend on the conserved charges $q_2,\ldots,q_L$ whose values satisfy the WKB quantization
conditions
\begin{align}\label{q-as}
\oint_{\alpha_k} dx \, p(x) = 2\pi\hbar (\ell_k+\ft12)\,,\qquad (\text{for $k=1,\ldots,L-1$})\,.
\end{align}
Here  integration goes over the cycles $\alpha_k$ on the complex curve \re{curve} encircling intervals on the real axis satisfying $y^2(x)>0$ and integers $\ell_k$ enumerate the 
quantized values of the charges $q_2,\ldots,q_L$ and the energy $\mathcal{E}_{N,L}=\mathcal{E}_{N,L}(\ell_1,\ldots,\ell_{L-2})$.
For large spin $N$ and twist $L$,   the minimal
energy $\mathcal{E}_{N,L}^{(0)}=\min_{\ell_k} \mathcal{E}_{N,L} $ has the following scaling behavior~\cite{Belitsky:2006en}
\begin{align}
\mathcal{E}_{N,L}^{(0)} = f(\rho)\ln N + O(N^0) \,,\qquad \rho=\frac{L}{\ln N}=\text{fixed}\,,\quad N,L\gg 1
\end{align}
where $f(\rho)$ is the so-called generalized scaling function. Detailed analysis of the relations \re{E-as} and \re{q-as} can be found in Refs.~\cite{Korchemsky:1995be,BDKM99,Belitsky00,GKK02,Belitsky:2006en}. For recent development
in the generalized scaling function in $\mathcal{N}=4$ SYM see review Ref.~\cite{chapter3}. 

So far we have discussed the exact solution for the one-loop anomalous dimensions of
quark and gluon maximal helicity operators. For the anomalous dimensions of mixed quark-gluon operators \re{O-mixed}, similar analysis of the spin chain 
\re{open}
can be carried out using Bethe Ansatz  for open $SL(2;\mathbb{R})$ spin chains~\cite{Belitsky00,DKM00}.

%One may wonder whether the QCD integrability phenomenon just described is preserved beyond one loop. We will address this question at the end of the next section.

\subsection{Integrability of dilatation operators in SYM theories}

In this subsection, we extend consideration to supersymmetric Yang-Mills theories.
Discussing integrability of dilatation operator in these theories, it is convenient to employ supersymmetric version of light-cone
formalism due to Mandelstam~\cite{Mandelstam83} and
Brink~{\it et al.}~\cite{BLN83}. In this formalism, all symmetries of SYM theory become manifest and
calculations can be performed in a unified manner for different
numbers of supercharges $\mathcal{N}=0,1,2,4$.
The maximally-supersymmetric $\mathcal{N}=4$ SYM theory is a finite,
four-dimensional conformal field theory~\cite{Mandelstam83,BLN83,SW81,HST84},
while the $\mathcal{N}=0$ theory corresponds to pure gluodynamics.

Defining a SYM theory on the light-cone, one starts with the
component form of the action, fixes the light-cone gauge $A_+(x) = 0$,  decompose all propagating, ``physical'' fields
into definite helicity components. In the case of $\mathcal{N}=4$ SYM, 
they include helicity $(\pm 1)$ fields, $A(x)$
and $\bar A(x)$, built from two-dimensional transverse
components of the gauge field,   complex scalar fields
$\phi^{AB}$ of helicity $0$ and helicity $\pm 1/2$ components of Majorana--Weyl
fermions, $\lambda^A$ and $\bar\lambda_A$, all in the adjoint representation of the
$SU(N_c)$ gauge group. An important
property of the light-cone formalism, which makes it advantageous over the
covariant one, is that the latter fields have only one non-vanishing component. As a consequence, one can describe helicity $(\pm
1/2)$ fermions by Grassmann-valued complex fields without any Lorentz index.
Introducing four fermionic coordinates $\theta^A$ (with $A=1,\ldots,4$) possessing the helicity $(-\ft12)$ and their conjugates $\bar\theta_A$ with helicity $\ft12$, we can assemble the above fields into a single, complex chiral $\mathcal{N}=4$ superfield~\cite{BLN83}
\begin{eqnarray}
\hspace{-3mm}\Phi (x, \theta^A) &=&  
\partial_+^{-1}A(x)
+ \theta^A \partial_+^{-1}\bar\lambda_A (x) + \frac{i}{2!} \theta^A \theta^B \bar
\phi_{AB} (x)
\nonumber\\
\hspace{-3mm}&-&\!\frac{1}{3!} \varepsilon_{ABCD} \theta^A \theta^B \theta^C \lambda^D
(x)\! - \frac{1}{4!} \varepsilon_{ABCD} \theta^A \theta^B \theta^C \theta^D
\partial_+ \bar{A} (x)   . \quad \ \label{N=4-field}
\end{eqnarray}
It embraces all particle helicities, from $-1$ to $1$ with half-integer step,
and, therefore, $\Phi (x, \theta^A)$ describes a CPT self-conjugate
supermultiplet.  

Gauge theories on the light-cone with less or no supersymmetry can be deduced
from the maximally supersymmetric $\mathcal{N}=4$ theory by removing ``unwanted''
physical fields. In the superfield formulation this amounts to a truncation of
the $\mathcal{N}=4$ superfield, or equivalently, reduction of the number of
fermionic directions in the superspace~\cite{BT83}.  For instance,
to get the ${\mathcal N} = 1$ superfields one removes three odd
coordinates $\theta^2=\theta^3 = \theta^4 = 0$, whereas for ${\mathcal N} =
0$ all $\theta$'s in \re{N=4-field} have to be set to zero.  Notice that under
this procedure the truncated $\mathcal{N}=2$, $\mathcal{N}=1$ and $\mathcal{N}=0$ theories involve only half of the fields described by the $\mathcal{N}-$extended
SYM theory and the other half of the needed particle content arises from the
complex conjugated superfields $\bar\Phi \equiv \Phi^\ast$. 
 Explicit expressions for the action of the SYM theory in terms of
the light-cone superfields can be found in Ref.~\cite{BDKM04}.

In a close analogy with \re{ggg}, 
we can introduce multiparticle single-trace operators built
from light-cone superfields
\begin{equation}
\label{WilsonOperators}
\mathbb{O}(Z_1,\ldots,Z_L) = {\rm tr} \, \{ \Phi (Z_1) \Phi (Z_2) \cdots \Phi
(Z_L) \} \,,
\end{equation}
where $\Phi(Z) \equiv \Phi^a(Z) t^a$ is a matrix ($SU(N_c)$) valued superfield and $Z=(x,\theta^A)$ denotes its position in
the superspace with four even coordinates, $x_\mu$, and $\mathcal{N}$ odd
coordinates, $\theta^A$   with $A=1,\ldots, \mathcal{N}$. In
addition, we choose all superfields to be located along the light-cone
direction in the four-dimensional Minkowski space  defined by the light-like
vector $n_\mu$ (with $n^2=0$), so that
$n \cdot A = A_+ =0$. Similarly to the QCD case, the positions
of the superfields on the light-cone are parameterized by real numbers
$x_\mu = z n_\mu$, 
$
\Phi(Z_k) \equiv \Phi(z_k n,\theta_k^A)\, .
$
The single-trace operators \re{WilsonOperators} represent a natural
generalization of nonlocal light-ray operators in QCD, cf. Eq.~\re{ggg}. To
obtain the latter it is sufficient to expand $\mathbb{O}(Z_1,\ldots,Z_L)$ in
powers of odd variables $\theta^{A_1}_1\ldots \theta^{A_L}_L$. As in QCD,
nonlocal operators \re{WilsonOperators} serve as generating functions for Wilson
operators with the maximal Lorentz spin and minimal twist equal to the number of
constituent fields $L$. Such operators define a representation of the %collinear 
$SL(2|\mathcal{N})$ subgroup of the full superconformal group.

Examining light-ray operators \re{WilsonOperators} in SYM theories with different number of supercharges, we find that $\mathcal{N}=4$ case is special. In $\mathcal{N}=4$ SYM theory there is only one independent chiral
superfield $\Phi(Z)$ and, as a consequence, the operators \re{WilsonOperators}
generate \textsl{all} Wilson operators of twist$-L$ built from $L$
fundamental fields. For $\mathcal{N}\le 2$, the superfield $\Phi(Z)$ and its conjugate
$\bar\Phi(Z)$ are independent of each other and, in addition to the operators in
\re{WilsonOperators}, one can introduce ``mixed'' operators built from both
superfields. This means that in the $\mathcal{N}=0,1$ and $2$ SYM theories, the
operators \re{WilsonOperators} only generate a certain subset of the existing
Wilson operators in the $SL(2|\mathcal{N})$ subsector. 

The light-ray operators \re{WilsonOperators} play a special role
as far as integrability is concerned. Namely, as was shown in Refs.~\cite{BDKM04}, the one-loop dilatation operator acting on the space of single-trace operators
\re{WilsonOperators} can be mapped in the multicolor limit into a Hamiltonian of
a completely integrable Heisenberg $SL(2|\mathcal{N})$ spin chain. As before,
the length of the spin chain coincides with the number of superfields in  \re{WilsonOperators} and spin operators are generators
of a collinear $SL(2|\mathcal{N})$ subgroup of the full superconformal
group~\cite{BDKM04}.

We recall that in SYM theories with  $\mathcal{N} \le 2$  supercharges the operators \re{WilsonOperators} only generate a subsector of Wilson operators
of twist $L$. To describe the remaining
operators, one has to consider single-trace operators built from both superfields, like $ {\rm tr} \, \{ \Phi (Z_1) \bar\Phi (Z_2) \cdots \Phi
(Z_L) \}$. 
For such operators, the one-loop dilatation operator involves 
the additional term describing the exchange interaction  between superfields on the light-cone  $\Phi\bar\Phi\to\bar\Phi\Phi$. It breaks integrability symmetry and generates a mass gap in the spectrum of the anomalous dimensions~\cite{BDKM99}. At the same time, for large values of the superconformal spin the exchange interaction vanishes
and integrability gets restored in the leading large spin asymptotics of the anomalous dimensions.

\subsection{Integrability in QCD and SYM beyond one loop}

It is well-known that the conformal symmetry is broken in QCD and SYM theories
with $\mathcal{N} < 4$ supercharges while in the maximally supersymmetric
$\mathcal{N}=4$ model it survives on the quantum level. However the conformal
anomaly modifies anomalous dimensions starting from two loops only and,
therefore, the one-loop dilatation operator inherits the conformal symmetry of
the classical theory~\cite{BKM03,Belitsky:2004cz}. 

Starting from two-loop order, the dilatation operator in
the $SL(2)$ sector acquires several new features. First, it receives conformal
symmetry breaking corrections arising both due to a nonzero beta-function and
a subtle symmetry-violating effect induced by the regularization procedure~\cite{Muller:1993hg}.  Second, the form of the
dilatation operator starts to depend on the representation of the fermion fields,
i.e., fundamental $SU(3)$ in QCD and adjoint $SU(N_c)$ in SYM theories. The
difference between the two is that it is only in the latter case that one can
select planar diagrams by going over to the multi-color limit, while in the
former case the large-$N_c$ counting is inapplicable and the two-loop dilatation
operator receives equally important contributions from both planar and nonplanar
Feynman graphs. Thus, by studying the two-loop dilatation operator in the $SL(2)$
sector we can identify what intrinsic properties of gauge theories (conformal symmetry, supersymmetry and/or planar limit) are responsible for the existence of the integrability phenomenon per se.

For an all-loop dilatation operator $\mathbb{H}
(\lambda)$, depending on 't Hooft coupling constant $\lambda=g^2 N_c/(8\pi^2)$ and acting on a
Wilson operator built from $L$ constituent fields and an arbitrary number of
covariant derivatives, integrability would require, in general, the existence of
$L$ conserved charges. Two of the charges---the light-cone component of the
total momentum of $L$ fields and the scaling dimension of the operator---follow
immediately from Lorentz covariance of the gauge theory. However, the identification of the remaining charges $q_k(\lambda)$ with
$k=3,\ldots,L$ is an extremely nontrivial task. The eigenvalues of the charges
$q_k$ define the complete set of quantum numbers parameterizing the eigenspectrum
of the dilatation operator. Integrability imposes a nontrivial analytical
structure of anomalous dimensions of Wilson operators and implies the double
degeneracy of eigenvalues with the opposite parity
\cite{Korchemsky95,GraMat95,BDKM99}. At the same time, breaking of integrability leads to lifting of the degeneracy
in the eigenspectrum of the one-loop dilatation operator.  

Explicit  two-loop calculation of the anomalous dimensions of the aforementioned
aligned-helicity fermionic operators in all SYM theories showed that 
the same relation between integrability and degeneracy of the eigenstates holds true to two loops. Namely, as was found in Refs.~\cite{Belitsky:2006av}, the desired pairing of eigenvalues occurs for three-gaugino operators in SYM
theories with $\mathcal{N}=1,2$ supercharges and the $SU(N_c)$ gauge group. 

The two-loops dilatation operator in SYM theories receives conformal
symmetry breaking contribution and, in addition, it depends on the
number of supercharges $\mathcal{N}$. The latter dependence comes about through the contribution of $2(\mathcal{N}-1)$ real scalars and $\mathcal{N}$ gaugino fields propagating inside loops. Both contributions to two-loop dilatation operator can be factored out (modulo an additive normalization factor) into a multiplicative c-number. This property makes the
eigenspectrum of the two-loop dilatation operator alike in all gauge theories
including the $\mathcal{N}=4$ SYM in which case the dilatation operator is believed to be integrable to all loops \cite{review}. Summarizing the results of two-loop calculations of the anomalous dimension in QCD and in SYM theories, integrability of the dilatation operator only requires the planar limit but it is sensitive neither to conformal symmetry, nor to supersymmetry~\cite{Belitsky:2006av}. For recent
discussion of  integrability in relation to non-planar corrections to the  anomalous dimensions in $\mathcal{N}=4$ SYM see review Ref.~\cite{chapter4}.

In this section, discussing the properties of anomalous dimensions 
we restricted ourselves to  the $SL(2)$ sector.
 There have been several  developments that we cannot address
here in detail. In particular,
an important observation was made in Ref.~\cite{Ferretti:2004ba}, where it was shown that the diagonal part of one-loop QCD evolution kernels governing the scale dependence of Wilson operators of arbitrary twist, can be written in a Hamiltonian form in terms of quadratic Casimir operators of the full conformal $SO(2,4)$ group. This observation was used in Ref.~\cite{Braun:2009vc} to
work out the non-diagonal parts of the evolution kernels for generic
twist-four operators.

\section{Integrability in high energy scattering}
 
In the previous section, we described how integrability emerges in the problem
of finding the dependence of the structure functions $F(x,Q^2)$ on the hard scale $Q^2$. In this section, we explain that yet another integrability symmetry arises in the high-energy limit. 

In application to the structure function $F(x,Q^2)$ this limit corresponds to  $x\to 0$ for fixed $Q^2$.
At small $x$, the invariant energy $s=Q^2(1-x)/x$ of colliding virtual photon and hadron becomes large and the structure function is expected to have Regge-like
scaling behavior $F(x,Q^2)\sim (1/x)^{\omega}$. In terms of moments \re{OPE},
this corresponds to appearance of the Regge pole at $N=\omega$ 
\begin{align}\label{lambda1}
\tilde F_N(q^2) = \int_0^1 dx\,x^{N-1} F(x,q^2) \sim \frac{1}{N-\omega}\,.
\end{align}
It is well-known \cite{BFKL} that perturbative corrections to $F(x,q^2)$ are enhanced at small $x$ by large logarithms $\sim (\alpha_s \ln(1/x))^p$. This raised
the hope that the Regge behavior \re{lambda1} can be derived in QCD from
 resummation of such corrections to all loops. Going to moments, the expansion  over $(\alpha_s\ln s)^p$ is traded for the expansion of $\tilde F_N(q^2)$ over $(\alpha_s/N)^p$.
 
\subsection{Evolution equation}
 
\begin{figure}[ht]
%\psfrag{1}[cc][cc]{$\scriptstyle 1$} \psfrag{2}[cc][cc]{$\scriptstyle 2$} \psfrag{N}[cc][cc]{$\scriptstyle L$}
%\psfrag{h}[cc][cc]{$h(p)$}  \psfrag{q}[cc][cc]{$\gamma^*(q)$}
%\centerline{{\epsfysize4cm \epsfbox{BKP.eps}}}
\ifarxiv
\centerline{\includegraphics[height=4cm]{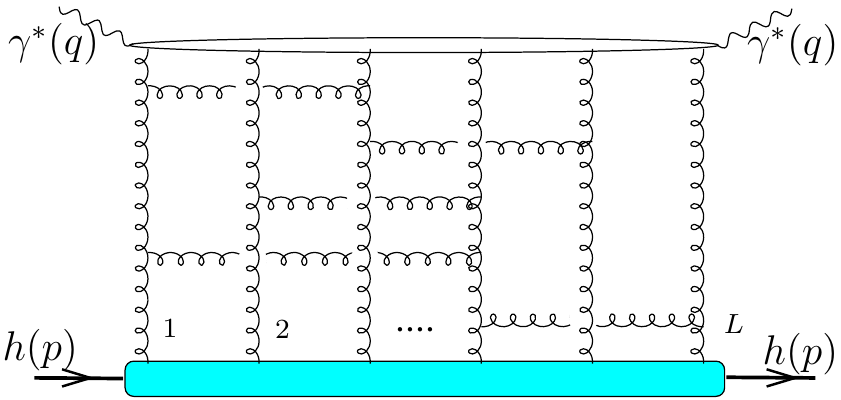}}
\else
\centerline{\includegraphics[height=4cm]{BKPtimes}}
\fi
\caption[]{\small The Feynman diagrams contributing to the deep inelastic scattering in the generalized leading logarithmic approximation. Wavy lines denote (reggeized) gluons. 
They couple to virtual photons through a quark loop.}
\label{diag}
\end{figure}

Careful study of asymptotic behavior of Feynman diagrams
describing interaction between virtual photon and hadron
shows that the dominant contribution to $F(x,q^2)$ only comes $t-$channel exchange of particles of maximal spin, i.e. gluons (see Fig.~2). Moreover, in the center-of-mass
frame of $\gamma^*(q)$ and $h(p)$, due to hierarchy of the scales, $s\gg Q^2$, interaction takes place in the two-dimensional plane orthogonal to the plane defined by the momenta of scattered particles, $p_\mu$ and $q_\mu$. This implies that in generic Yang-Mills theory the leading high-energy asymptotic behavior of the scattering amplitudes is driven by $t-$channel exchange of an arbitrary number of gluons. In the so-called generalized leading logarithmic approximation,
their contribution to the moments \re{lambda1} takes the form
\begin{align}\label{F-T}
\tilde F_N(q^2) = \sum_{L\ge 2} \int [d^2k]\int [d^2k'] \ \Phi_{\gamma^*}(\{k\}) \,T_L(\{k\},\{k'\};N)\, \Phi_{h}(\{k'\})\,,
\end{align}
where integration goes over two-dimensional momenta of $L$ gluons propagating
in the $t-$channel, $[d^2k]=\prod_1^L d^2k_i$ and similarly for $[d^2 k']$. Here, the wave functions $\Phi_{\gamma^*}(\{k\})$ and  $\Phi_{h}(\{k'\})$ describe the coupling of $L$ gluons to virtual photon and hadron, respectively. Also, $T_L(\{k\},\{k'\};N)$ describes 
elastic  scattering of $L$ gluons in the $t-$chan\-nel  (see Fig.~\ref{diag}) and is the main object of 
our consideration. 

It is convenient to rewrite \re{F-T} as the following matrix element
\begin{align}\label{F-vev}
\tilde F_N(q^2) = \sum_{L\ge 2} \vev{\Phi_{\gamma^*}|T_L(N)|\Phi_{h}}\,,
\end{align}
where  the minimum number of two gluons, $L=2$, is required in order to get a colorless exchange. The transition operator $T_L(N)$ describes the elastic
scattering of $L$ gluons. In the generalized leading logarithmic approximation,
the Feynman diagrams contribution to $T_L(N)$ have ladder structure as shown in Fig~\ref{diag}. They can be
resummed leading to the following Bethe-Salpeter  equation~\cite{Bartels80,KP80}
\begin{align}\label{BS}
N T_L(N) = T_L^{(0)} + \frac{\alpha_s}{2\pi} \mathbb{H}_L\,T_L(N)\,,
\end{align}
where $ T_L^{(0)}$ corresponds to the free propagation of $L$ gluons in the $t-$channel and the evolution operator $\mathbb{H}_L$ describes their pair-wise interaction. The operator $\mathbb{H}_L$ acts both  on two-dimensional momenta
and  on colors of $L$ gluons and  has the following two-particle form
\begin{align}
\mathbb{H}_L = \sum_{1\le i<j\le L} H_{ij}\, t_i^a t_j^a\,.
\end{align}
Each term in this sum is given by the product of the color factor involving color
charges of two gluons and two-particle kernel $H_{ij}$ acting locally on the tranverse momenta of gluons with indices $i$ and $j$. The kernel $H_{ij}$ is known as BFKL
operator \cite{BFKL} and it is defined below in \re{H-oper}. 

Combining together \re{BS} and \re{F-vev} we obtain the following expression for $\tilde F_N(q^2)$
\begin{align}
\tilde F_N(q^2) = \sum_{L\ge 2} \vev{\Phi_{\gamma^*}|\lr{N- \frac{\alpha_s}{2\pi} \mathbb{H}_L}^{-1} T_L^{(0)}|\Phi_{h}}\,.
\end{align}
We observe that $\tilde F_N(q^2)$ has (Regge) singularities in $N$ which are determined by the eigenspectrum of the operator $\mathbb{H}_n$, the so-called
BKP equation \cite{Bartels80,KP80},
\begin{equation}
{\mathbb H}_L \Psi_{L,\{q\}}(k_1,\ldots,k_L) = E_{L,\{q\}}
\Psi_{L,\{q\}}(k_1,\ldots,k_L) \,.
\label{BKP}
\end{equation}
The solutions to \re{BKP} define color singlet compound states
of $L$ gluons and we introduced $\{q\}$ to denote the set of quantum numbers parameterizing all solutions. Having solved Schr\"odinger like equation \re{BKP},
we can compute the moments of the structure function as~\cite{Korchemsky95}
\begin{align}\label{sol}
\tilde F_N(q^2) = \sum_{L\ge 2} \sum_{\{q\}}  \lr{N- \frac{\alpha_s}{2\pi} E_{L,\{q\}}}^{-1}  \beta_{L,\{q\}}\,.
\end{align}
Here the impact factor $\beta_{L,\{q\}}= \vev{\Phi_{\gamma^*}|\Psi_{L,\{q\}}}\vev{\Psi_{L,\{q\}}|T_L^{(0)}|\Phi_{h}}$ measures the projection of the eigenstates onto the wave functions
of scattered particles. The double sum in \re{sol} runs over the possible number of gluons $L\ge 2$ and
over all eigenstates of the BKP Hamiltonian \re{BKP} parameterized by the conserved 
charges $q$.
We observe that this relation has an expected Regge form \re{lambda1}.
Moreover, the leading Regge behavior of the structure function is controlled
by right-most singularity of  $\tilde F_N(q^2)$ in complex $N$ plane. According to \re{sol}, it corresponds to  the {\it maximal} value of the `energy' $E_{L,\{q\}}$.

\subsection{Conformal \texorpdfstring{$SL(2;\mathbb{C})$}{SL(2;C)} symmetry}

We recall that $k_i$ in the BKP equation \re{BKP} describe two-dimensional transverse momenta of $i$th gluon and the relation \re{BKP} can be interpreted as
two-dimensional Schr\"o\-din\-ger equation for $n$ particles carrying $SU(N_c)$ color charges. 

As was found in \cite{Lipatov90,Lipatov94,FK95}, the BKP equation \re{BKP} becomes integrable in the multi-color limit. In this limit, the relevant ladder Feynman diagrams contributing to $\tilde F_N(q^2)$ have the topology of a cylinder and, as a consequence, the evolution operator ${\mathbb H}_L$ reduces to the sum of terms corresponding to
pairwise nearest-neighbor BFKL interactions:
\begin{equation}
{\mathbb H}_L = \frac12 \sum_{k=1}^L { H}_{k,k+1} + O(1/N_c^2)
\label{H-multi}
\end{equation}
with periodic boundary conditions ${H}_{L,L+1}={H}_{L,1}$. Notice that this
relation is exact for $L=2$.

The BFKL operator ${H}_{k,k+1}$ has a number of remarkable properties
which allow us to solve the Schr\"odinger equation (\ref{BKP})
exactly~\cite{Lipatov85,Lipatov90}. To elucidate these properties it is convenient
to switch from two-dimensional momenta $k_i$ to two-dimensional coordinates $b_i$ via Fourier transform and, then,  introduce complex  holomorphic and the
antiholomorphic coordinates
\begin{equation}
\vec{k}_i\quad \mapsto\quad  \vec{b}_i=\{x_i,y_i\} \quad \mapsto \quad   (z_{i} = x_{i}+i
y_{i}\,,\quad \bar{z}_{i} = x_{i}-i y_{i})\,.
\end{equation}
Quite remarkably,  ${H}_{12}$ is invariant under the conformal $SL(2;\mathbb{C})$ transformations of the gluon coordinates on the plane~\cite{Lipatov85,Lipatov90}
\begin{equation}
z_k \to \frac{a z_k + b }{c z_k + d}\,,\qquad (ad-bc=1)\,,
\end{equation}
and similarly for antiholomorphic coordinates $\bar z_k$.
The generators of these transformations are
\begin{equation}\label{c-gen}
L_{k,-}=-\partial_{z_k}\,,\qquad L_{k,0}=z_k\partial_{z_k}\,,\qquad
L_{k,+}=z_k^2\partial_{z_k}\,,
\end{equation}
and the corresponding antiholomorphic  generators $\bar L_{k,-}$, $\bar
L_{k,0}$ and $\bar L_{k,+}$ are given by similar expressions with $z_k$ replaced
by $\bar z_k$, with $k=1,2$ enumerating particles. Then, ${H}_{12}$ commutes with all two-particle generators
\begin{equation}\label{?-1}
[ {H}_{12},L_{1,a}+L_{2,a}] =
[ {H}_{12},\bar L_{1,a}+\bar L_{2,a}] = 0\,
\end{equation}
with $a=+,-,0$. This implies that, firstly,  ${H}_{12}$ only depends on the
two-particle Casimir operators of the $SL(2,\mathbb{C})$ group
\begin{equation}\label{SL2C-Casimir}
{L}_{12}^2=-(z_1-z_2)^2\partial_{z_1}\partial_{z_2}\,,\qquad
\bar{{L}}_{12}^2=-(\bar z_1-\bar z_2)^2\partial_{\bar z_1}
\partial_{\bar z_2}\,,
\end{equation}
and, secondly, the eigenstates of ${H}_{12}$ have 
to diagonalize the Casimir
operators
\begin{equation}\label{eig-SL2C}
{L}_{12}^2\Psi_{n,\nu}=h(h-1) \Psi_{n,\nu}\,,\qquad
\bar {L}_{12}^2\Psi_{n,\nu}=\bar h(\bar h-1) \Psi_{n,\nu}\, .
\end{equation}
Here a pair of complex conformal spins is introduced
\begin{equation}\label{h}
h=\frac{1+n}2+i\nu\,, \qquad \quad
\bar h=\frac{1-n}2+i\nu\,
\end{equation}
with a non-negative integer $n$ and real $\nu$ that specify the irreducible
(principal series) representation of the $SL(2,\mathbb{C})$ group to which
$\Psi_{n,\nu}$ belongs. The solutions to Eqs.\ (\ref{eig-SL2C}) are \cite{Lipatov85}
\begin{equation}\label{wf-2}
\Psi_{n,\nu}(b_1,b_2) =\left(\frac{z_{12}}{z_{10}z_{20}}\right)^{(1+n)/2+i\nu}
\left(\frac{\bar z_{12}}{\bar z_{10}\bar z_{20}}\right)^{(1-n)/2+i\nu}\,,
\end{equation}
where $z_{jk}=z_j-z_k$ and $b_0=(z_0,\bar z_0)$ is the collective
coordinate, reflecting the invariance of $ {H}_{12}$ under
translations. The corresponding eigenvalue of $H_{12}$ reads~\cite{BFKL,Lipatov85}
\begin{equation}\label{2-en}
E_{n,\nu}=2\psi(1)-\psi\left(\frac{n+1}2+i\nu\right)
-\psi\left(\frac{n+1}2-i\nu\right)\,.
\end{equation}
Its maximal value, $\textrm{max}\,E_{n,\nu}=4\ln 2$, corresponds to $n=\nu=0$,
or equivalently $h=\bar h=1/2$. It defines the position of the right-most
singularity $\omega=4\ln 2 \,{\alpha_s N_c}/{\pi} $ in \re{lambda1} 
known as the BFKL pomeron~\cite{BFKL}.
The relations \re{wf-2} and \re{2-en} define the exact solution to the
Schr\"odinger equation \re{BKP} for $n=2$, that is for the color-singlet compound state built from two reggeized gluons. 

\subsection{Heisenberg \texorpdfstring{$SL(2;\mathbb{C})$}{SL(2;C)} spin chain} 
 
Using (\ref{2-en}) one can reconstruct the operator form of the BFKL kernel
${H}_{12}$ on the representation space of the principal series of
the $SL(2,\mathbb{C})$ group
\begin{equation}\label{H-oper}
{H}_{12}=\frac12\left[H(J_{12}) + H(\bar J_{12})\right] \,,\qquad
H(J)=2\psi(1)-\psi(J)-\psi(1-J)\,,
\end{equation}
where, as before, the two-particle spins are defined as ${L}_{12}^2 = J_{12}
(J_{12}-1)$ and $\bar{L}_{12}^2=\bar J_{12}(\bar J_{12}-1)$. Notice that we
already encountered the similar Hamiltonian in Sect.~2 (see Eq.~\re{closed}) and found that it gives rise to integrability for the dilatation operator.

Most remarkably, the Hamiltonian \re{H-multi}  has the same hidden integrability as
the dilatation operator \re{closed} and it coincides in fact with the  Hamiltonian of
the $SL(2,\mathbb{C})$ Heisenberg magnet~\cite{Lipatov94,FK95}. 
The important difference
between the two operators is that they are defined on the different space
of functions: the operator \re{closed}  acts on the nonlocal light-ray operators \re{O-def} which are polynomials in the light-cone coordinates while the eigenfunctions of the operator
\re{H-oper} are single-valued functions on the two-dimensional plane, Eq.~\re{wf-2}. 
This leads to a dramatic change in the properties of the two evolution kernels.
 
The number of sites in the Heisenberg $SL(2,\mathbb{C})$ spin chain \re{H-multi}
equals the number of particles and the corresponding spin operators are
identified as six generators, $L_k^\pm, L_k^0$ and $\bar L_k^\pm,\bar L_k^0$,
of the $SL(2,\mathbb{C})$ group.   It possesses a large-enough set
of mutually commuting conserved charges $q_n$ and $\bar q_n$ $(n=2,...,L)$ such
that $\bar q_n= q_n^\dagger$ and $[\mathbb{H}_L,q_n]=[\mathbb{H}_L,\bar
q_n]=0$. The charges $q_n$ are polynomials of degree $n$ in the holomorphic
spin operators. 
They have the following form~\cite{Lipatov94,FK95}
\begin{equation}
q_n= \sum_{1\le j_1 < j_2 < ... < j_n \le L} z_{j_1j_2}z_{j_2j_3}... z_{j_nj_1}
p_{j_1}p_{j_2}...p_{j_n}
\label{q_n}
\end{equation}
with $z_{jk}=z_j-z_k$ and $p_j=i\partial_{z_j}$. The ``lowest'' charge $q_2$ is
related to the total spin of the system $h$. For the principal series of the
$SL(2,\mathbb{C})$ it takes the following values
\begin{equation}
q_2=-h(h-1)\,,\qquad h=\frac{1+n_h}2+i\nu_h\,,
\label{h1}
\end{equation}
with $n_h$ integer and $\nu_h$ real. The eigenvalues of the integrals of motion,
$q_2,...,q_L$, form the complete set of quantum numbers parameterizing the
$L-$gluon states \re{BKP}.

Identification of \re{H-multi} as the Hamiltonian of the $SL(2,\mathbb{C})$
Heisenberg magnet allows us to map the $L-$gluon states into the eigenstates
of this lattice model. In spite of the fact that the Heisenberg $SL(2,\mathbb{C})$  magnet represents a generalization of the $SL(2,\mathbb{R})$ spin chain, finding its exact solution
is a much more complicated task. The principal
difficulty is that, in distinction with $SL(2,\mathbb{R})$  magnet, the quantum space of
the $SL(2,\mathbb{C})$ magnet does not possess a highest weight -- the so-called
``pseudo-vacuum state'' -- and, as a consequence, conventional methods like
the Algebraic Bethe Ansatz method~\cite{QISM} are not applicable. The eigenproblem
\re{H-multi} has been solved exactly in Refs.~\cite{DKM01,KKM02,DKKM02} using the
method of the Baxter ${Q}-$operator~\cite{Baxter,Sklyanin,Bazhanov:1989nc,PG92} which does not rely on
the existence of a  highest weight. In this approach, it becomes possible to
establish the quantization conditions for the integrals of motion $q_3,...,q_L$ and
to obtain an explicit form for the dependence of the energy $E_L$ on the integrals
of motion.

In this manner, the spectrum of the $L-$gluon state has been calculated for $L\ge 3$ particles: For $L=3$ few low-lying states have been found in
\cite{JW99,BLV99} and the complete spectrum of states for $3 \le L \le 8$ was
determined in \cite{KKM02,DKKM02} (see also \cite{VL01}). The obtained eigenspectrum
has a very rich structure. The quantized values of the conserved $q-$charges and the energy $E_L$ depend on the integer $n_h$ and the real number $\nu_h$ defining the total $SL(2,\mathbb{C})$ spin of the state, Eq.~\re{h}. In addition, they also depend on the ``hidden'' set of integers
$\Mybf{\ell}=(\ell_1,\ell_2,...,\ell_{2(L-2)})$. As a function of $\nu_h$, the charges form a
family of trajectories in the moduli space $\Mybf{q}=(q_2,q_3,...,q_L)$ labelled
by integers $n_h$ and $\Mybf{\ell}$. Each trajectory in the $q-$space induces
a corresponding trajectory for the energy $E_L=E_L(\nu_h;n_h,\Mybf{\ell})$.
The origin of these trajectories and the physical interpretation of the integers
$\Mybf{\ell}$  can be understood by solving the Schr\"odinger
equation \re{BKP} within the semiclassical approach described in the next subsection.
 
\subsection{Semiclassical limit}

In the semiclassical approach \cite{DKM03}, we assume that the $SL(2;\mathbb{C})$ spins $h$ and $\bar h$
are large and apply the WKB methods to construct the asymptotic solution to \re{BKP}. 
One might expect a priori  that this approach could be applicable only for
highly-excited states. Nevertheless, as was demonstrated in \cite{DKM03}, the
semi-classical formulae work with good accuracy throughout the whole spectrum.

{}From the viewpoint of classical dynamics, the multi-gluon states \re{BKP} are describe 
by a chain of interacting particles `living' on the two-dimensional $\vec b-$plane
\cite{Korchemsky:1997yy,GKK02}. The classical model inherits the complete
integrability of the quantum noncompact spin magnet. Its Hamiltonian and the
integrals of motion are obtained from \re{H-multi}, \re{H-oper} and \re{q_n}
by replacing the momentum operators by the corresponding classical functions.
Since the Hamiltonian \re{H-multi} is given by the sum of holomorphic and
antiholomorphic functions, from point of view of classical dynamics the model
describes two copies of one-dimensional systems defined on the complex $z-$
and $\bar z-$lines. The solutions to the classical equations of motion have a
rich structure and turn out to be intrinsically related to the finite-gap
solutions to the nonlinear equations \cite{NMPZ84,Krichever77}; namely, the
classical trajectories have the form of plane waves propagating in the chain
of $L$ particles. Their explicit form in terms of the Riemann $\theta-$functions
was established in \cite{Korchemsky:1997yy} by the methods of finite-gap theory
\cite{NMPZ84,Krichever77}. 
 
In the semiclassical approach, the eigenfunctions in
\re{H-multi} have the standard WKB form, $\Psi_{\rm WKB}(\vec z_1, \ldots,
\vec z_L)\sim \exp(iS_0/\hbar)$ where the Planck constant $\hbar = |q_2|^{-1/2}$ is related to the lowest charge \re{h1} and the action function $S_0$ satisfies the Hamilton-Jacobi
equations in the classical $SL(2;\mathbb{C})$ spin chain. It turns out that the solutions to 
these equations are determined by the same spectral curve \re{curve} 
as for the $SL(2;\mathbb{R})$ spin chain. The charges $\Mybf{q}$ define the moduli of this curve and take arbitrary complex values in the classical model.
Going over to the quantum model, we find  that charges $\Mybf{q}$ are quantized.

The quantization
conditions for the charges $\Mybf{q}$ follow from the requirement that
$\Psi_{\rm WKB}(\vec z_1, \ldots,\vec z_L)$ has to be a single-valued function of $\vec z_i$. As was shown in
Refs.~\cite{GKK02,DKM03}, these conditions can be expressed in terms of the
periods of the ``action'' differential over the canonical set of the $\alpha-$
and $\beta-$cycles on the Riemann surface corresponding to the complex curve
\re{curve}
\begin{equation}
\Re  \oint_{\alpha_k}dx\,p(x) =\pi\,\ell_{2k-1}
\, , \qquad
\Re  \oint_{\beta_k}dx\,p(x) =\pi \, \ell_{2k}
\, ,
\label{WKB-intro}
\end{equation}
with $k=1,\ldots,L-2$ and $\Mybf{\ell}=(\ell_1,\ldots,\ell_{2L-4})$ being the set
of integers. The relations \re{WKB-intro} define the system of $2(L-2)$ real
equations for $(L-2)$ complex charges $q_3,...,q_L$ (we recall that the
eigenvalues of the ``lowest'' charge $q_2$ are given by \re{h1}). Their
solution leads to the semiclassical expression for the eigenvalues of the
conserved charges. In turn, the energy of the $L-$gluon states $E_{L,q}$
can be expressed  as a function of $q_3,...,q_L$. In the semiclassical approach,
the corresponding expression is
\begin{eqnarray}\label{E-specc}
E_L^{\rm (as)}
&=& 4 \ln 2
\\
&+& 2 \Re  \sum_{k=0}^L
\bigg[
\psi(1 + i \Re  \delta_k + | \Im \delta_k|)
+
\psi(i \Re \delta_k + |\Im \delta_k|) - 2 \psi(1)
\bigg]
\, , \nonumber
\end{eqnarray}
where $\delta_k$ are roots of the polynomial $t_L(u)$ defined in  \re{transfer}.

{The expression in Eq.~\re{E-specc} is similar to the energy of
the $SL(2,\mathbb{R})$ magnet in Eq.~\re{E-as} although the properties
of the two models are different.} As was demonstrated in
\cite{DKM03}, the resulting semiclassical expressions for $q_3,...,q_L$ and
$E_L$ are in good agreement with exact results~\cite{KKM02,DKKM02}.
A novel feature of the quantization conditions \re{WKB-intro} is that they
involve \textsl{both} the $\alpha-$ and $\beta-$periods on the Riemann surface.
This should be compared with the situation in the Heisenberg $SL(2,\mathbb{R})$
magnet discussed in Section~2.6. There, the WKB quantization conditions involve
only the $\alpha-$cycles, Eq.~\re{q-as}, since the $\beta-$cycles correspond to
classically forbidden zones. For the $SL(2,\mathbb{C})$  magnet, the classical
trajectories wrap over an arbitrary closed contour on the spectral curve
\re{curve} leading to \re{WKB-intro}. This fact allows us to explore the full
modular group~\cite{Dubrovin81} of the complex curve \re{curve} and explain
the fine structure of the exact eigenspectrum of the $SL(2;\mathbb{C})$ magnet.
More details can be found in  Ref.~\cite{DKM03}.

 \section{Concluding remarks}
 
In this review, we have described integrability symmetry in application to the deeply inelastic scattering in QCD.  Due to space limitations, we did not review various important topics
and  we refer the interested reader to Ref.~\cite{Belitsky:2004cz} for a comprehensive review on the subject.
We would like to emphasize that integrability is not of a mere academic interest in QCD as it offers a powerful technique for solving important phenomenological problems such as  finding  the scale dependence of hadronic structure functions of higher twist and describing their high-energy (Regge) asymptotic behaviour.  On theory side, the very fact that QCD
evolution equations exhibit integrability property provides yet another indication that QCD
possesses some hidden (integrable) structures waiting to be uncovered.

\section*{Acknowledgements}

It is a pleasure to thank A.~Belitsky, V.~Braun, S.~Derkachov, A.~Gorsky, A.~Manashov
and D.~M\"uller for an enjoyable collaboration on the topics reviewed above.

\phantomsection

\end{document}